\documentclass[12pt]{iopart}

\usepackage{amssymb}
\usepackage{iopams}
\usepackage{graphicx}

\newcommand \be {\begin{equation}}
\newcommand \ee {\end{equation}}
\newcommand \bea {\begin{eqnarray}}
\newcommand \eea {\end{eqnarray}}
  
\begin{document}

\bibliographystyle{unsrt}

\title[Free volume distributions and compactivity in granular media]
{Free volume distributions and compactivity measurement in a bidimensional
granular packing}

\author{F Lechenault$^1$, F da Cruz$^1$, O Dauchot$^1$ and E Bertin$^2$}

\address{$^1$ Commissariat \`a l'Energie Atomique, F-91191 Saclay, France}
\address{$^2$ Department of Theoretical Physics, University of Geneva,
CH-1211 Geneva 4, Switzerland}
\ead{olivier.dauchot@cea.fr}

\begin{abstract}
We investigate experimentally the distribution of the free volume inside a
bidimensional granular packing associated either to single grains or to
clusters of grains. This is done for two different kinds of grains and two
levels of compaction. The logarithm of the free volume distribution scales in
a nonextensive way with the cluster size for cluster sizes up to a few hundred
grains. Having factorized this size dependence, we characterize the
distributions by two intensive parameters. We discuss  the interpretation of
these parameters and their possible relation to Edwards' compactivity.
\end{abstract}

\section{Introduction}
Granular media consist in a large number of grains and one would like to have
general predictions for the macroscopic laws ruling such media without having 
to deal with individual components. Recent experiments suggest that such a
thermodynamic description might be feasible~\cite{Nowak98,Schroter05}, and
there is a strong motivation for providing a statistical ground to this
hypothetic description
\cite{Edwards89,Barrat00,Marty05,Schroter05,Dauchot05,Makse02,Dean03}.
Yet, a theoretical understanding is still lacking regarding the way
microscopic configurations are explored dynamically.

It has been argued by Edwards and collaborators~\cite{Edwards89,Mehta89}
that the dynamics is controlled by the mechanically stable --the so-called
blocked-- configurations and that all such configurations of a given volume
are statistically equivalent provided that the driving involves extensive
manipulations, such as shaking, shearing or pouring. This immediately leads
to the definition of a configurational entropy $S(V)$ which counts the
number of such configurations compatible with a given volume. The
associated state variable, the ``compactivity'' $X$, then naturally follows
from the relation $X^{-1}=\partial S/\partial V$.
So far, this equiprobability assumption has been tested through numerical
simulations in different physical contexts: in schematic
finite-dimensional models with kinetic constraints~\cite{Barrat00}, in
spin glass models with non-thermal driving between the blocked
states~\cite{Dean01} and finally in a few more realistic models of particle
deposition~\cite{Brey00} or molecular dynamics simulations of
granular media driven by an external shear~\cite{Makse02}.
Note however that Edwards' approach has also been shown to fail in some
cases~\cite{DeSmedt02,Godreche05}.

>From an experimental point of view, a stationary dynamics has been observed
for vertically vibrated glass beads at rather high volume fraction, and has
been
shown to be independent of the history of the system~\cite{Nowak98}.
In addition, a relation between the fluctuations of volume fraction and
the volume fraction itself has recently been reported~\cite{Schroter05}
in a similar experimental system. Assuming that usual thermodynamic relations
are valid, this led to a determination of the compactivity $X$ of the packing.
An alternative determination of $X$ has also been proposed from the
exponential decay of single grain volume distribution~\cite{Richard03,Aste05b}.

In spite of these globally encouraging results, the proposal made by Edwards
and coworkers still needs to be clarified in several respects.
First, the choice of the physical variables to be used in
practical situations essentially remains an open issue~\cite{Bouchaud02},
and the computation of the corresponding density of state turns out to
be out of reach of present theories~\cite{Blumenfeld05}.
Second, this statistical construction may be strongly affected by the
presence of correlations that extend over length scales comparable to the
system size. Indeed the occurrence of such long-range correlations has been
reported in various contexts, for instance in jammed granular systems
\cite{Donev05,Somfai05}, in porous media \cite{Makse96}, or in cellular
structures \cite{Godreche92}. In addition, recent theoretical advances
suggest that amorphous systems are close to a critical point associated with
a jamming transition \cite{Hern02,Hern03,Wyart05}, thus also supporting
the existence of correlations on length scales small than a correlation
length that may be rather large. These various issues clearly call for
experimental checks of the foundation of Edwards' approach.

In this paper, we report on an experimental investigation of the
distributions of volume, as well as free volume, of Voronoi cells associated
with each grains in a bidimensional granular packing. The distribution
$P(v_N^f)$ of the free volume per grain $v_N^f$ in clusters of neighboring
grains is also investigated in a systematic way as a function of the size $N$
of the clusters. Two different types of grains, as well as two different
levels of compaction, are successively used. Our central result is that
$\ln P(v_N^f)$ scales in a nonextensive way as $N^{\alpha} g(v^f_N,\chi_e,\eta_e)$
in the accessible range of $N$, with $\alpha \approx 0.8$.
The parameter $\chi_e$ and $\eta_e$ are intensive, and $\chi_e$ may be
thought of as a generalization of the compactivity introduced by Edwards.

The paper is organized as follows. In Sect.~2, we briefly recast
Edwards' formalism into a more general form which allows possible
departures from the equiprobability of states to be taken into account.
The experimental set-up is presented in Sect.~3, and the experimental
results on the distributions of volume and free volume are reported
in Sect.~4. Finally, Sect.~5 is devoted to a discussion of the physical
interpretation and consequences of our results.

\section{General theoretical formulation}
In order to propose an experimental procedure for the determination of
statistical properties of granular packings, we first discuss in more
details the theoretical background.
Although the definition of compactivity apparently relies on the
assumption that the states compatible with a given volume are equiprobable,
a compactivity can actually be defined in a more general context.
Let us consider a granular packing, and denote its microscopic
configurations by $\mathcal{C}$. In the spirit of Edwards, we assume that
the volume is a globally conserved quantity. On general grounds, the
steady-state probability $P(\mathcal{C})$ may be written as, taking into
account the conservation of volume
\be \label{dist-mucan}
P(\mathcal{C}) = \frac{f(\mathcal{C})}{Z(V)}\, \delta(V(\mathcal{C})-V)
\ee
which defines the probability weight $f(\mathcal{C})$ and the
normalization factor $Z(V)$. Let us assume that, on large scale, different
regions of the system are only correlated by the global volume conservation. By this we mean that, considering a partition of the system into two
subsystems $\mathcal{S}_{1,2}$, $Z(V)$ factorizes into $Z(V) \approx
Z_1(V_1)\, Z_2(V-V_1)$ --the approximate equality results from possible
interface contributions, that are expected to vanish in the thermodynamic
limit. This is in particular the case if the probability weights
$f(\mathcal{C})$ are assumed to factorize as
$f(\mathcal{C}) = f_1(\mathcal{C}_1)f_2(\mathcal{C}_2)$, where
$\mathcal{C}=(\mathcal{C}_1,\mathcal{C}_2)$. Under such assumptions,
a compactivity can be defined as a generalized "intensive thermodynamic
parameter" for out-of-equilibrium systems~\cite{Bertin06}:
\be \label{compactivity}
\frac{1}{X}=\frac{\partial \ln Z}{\partial V}.
\ee
Note that Edwards' definition is recovered when $f(\mathcal{C})$ is uniform
over blocked configuration, and zero for other configurations.
This reformulation shows that the equiprobability assumption
is actually unnecessary to define a compactivity.

A relevant experimental situation consists in studying a small part of a large
packing. In this case, the effect of the rest of the system, acting as a
reservoir of volume, may be accounted for by its compactivity. Integrating out
the degrees of freedom of the reservoir in distribution~(\ref{dist-mucan})
yields the ``canonical'' distribution for the subsystem
\be \label{dist-can}
P_c(\mathcal{C}) = \frac{f(\mathcal{C})}{Z_c(X)}\,
e^{-V(\mathcal{C})/X}
\ee
Interestingly, due to the presence of the exponential factor in
Eq.~(\ref{dist-can}), usual thermodynamic relations like
\be
\langle \delta V^2\rangle = X^2 \frac{\partial V}{\partial X}
\ee
are still valid independently of the specific form of the probability weights
$f(\mathcal{C})$, provided that the above factorization assumption is
satisfied. Such thermodynamic relations offer the possibility to measure
the compactivity of a given packing, as done in~\cite{Schroter05}.
Yet, more information may be gained by measuring the probability of observing
a volume $V$ in a subsystem of $N$ grains. This distribution $\Phi_N(V)$ is
obtained by summing the canonical distribution (\ref{dist-can}) over all
configurations with a given volume $V$:
\be \label{V_dist}
\Phi_N(V) = \int d\mathcal{C} P_c(\mathcal{C})\, \delta(V(\mathcal{C})-V) =
\frac{Z_N(V)}{Z_c(X)}\, e^{-V/X} = \frac{1}{Z_c(X)}\, e^{S_N(V)-V/X} 
\ee
Apart from the exponential factor, most of the information about the system
lies in the function $Z_N(V)$, the logarithm of which is a generalized
entropy $S_N(V)=\ln Z_N(V)$. Computing $S_N(V)$, which depends in a
complicated way on the (usually unknown) probability weights $f(\mathcal{C})$
is a formidable task, out of reach of present theories. Hence, it is of
primary importance to gather quantitative informations on $S_N(V)$ and its
scaling in $N$ from the experimental side. 

To this aim, one would first need to choose a set of physical variables
that are both experimentally accessible, and relevant to describe
a microscopic configuration. These variables may actually not fully define
the configuration, as different configurations might lead to the same values
of the variables.
The most simple choice is to consider the volumes $v_i$ of cells associated
to each grain through some tiling of space~\cite{Blumenfeld02}, leaving
aside from the description the forces between grains.
Alternatively, the free volumes $v^f_i$, defined by subtracting to $v_i$
the volume of the grain, may also be a relevant description, more convenient
for polydisperse systems. In this paper, we consider both descriptions,
with still an emphasis on the free volume description.
In addition, we focus on systems, like our experimental set-up presented
below, that are constrained to evolve among blocked states.
Such systems may then be described by a density of state $\rho(\{v^f_i\})$,
which corresponds to the number of blocked states with given free volumes
$v^f_i$. Introducing the total free volume of the packing $V^f=V-V_g$,
with $V_g$ the total volume occupied by the grains, the function $Z_N(V^f)$
reads
\be \label{auboisdormant}
Z_N(V^f)=\int\prod_i dv^f_i\, \rho(\{v^f_i\})\, f(\{v^f_i\})\,
\delta\left(\sum_i v^f_i - V^f\right)
\ee
Let us comment briefly on the physical interpretation of the above equation.
First, we note that $\rho(\{v^f_i\})$ is the static density of states,
which corresponds to a simple ``counting'' of blocked states compatible
with the constraints. $\rho(\{v^f_i\})$ encodes a compatibility condition regarding the  variables that are not taken into account in the chosen microscopic description - for example the forces - and hence measures the degeneracy of the corresponding configuration. Second, the dynamical population of the states is accounted for by the dynamical probability weight $f(\{v^f_i\})$.
This factor precisely encodes the possible departure from equiprobability.
Thus these two terms clearly have a different interpretation.
Yet, only their product appears in equation~(\ref{auboisdormant}), so that
it is not possible to determine $f(\{v^f_i\})$ without an
independent knowledge of $\rho(\{v^f_i\})$. As such a determination
of $\rho(\{v^f_i\})$ is in general out of reach in a realistic experimental
system, it follows that the validity of Edwards' assumption on the
equiprobability of states cannot be tested experimentally.
Only in some simplified theoretical models $\rho(\{v^f_i\})$ may be
determined, and the validity of Edwards' assumption may be tested
explicitly~\cite{Barrat00,Makse02,Godreche05}.
Yet, as explained above, Edwards' assumption is not necessary to implement
the above statistical formulation and to motivate experimental determination
of the (generalized) entropy and of the associated compactivity.

\section{Experimental set-up}

\begin{figure}[t]
\centering\includegraphics[width=8cm,clip]{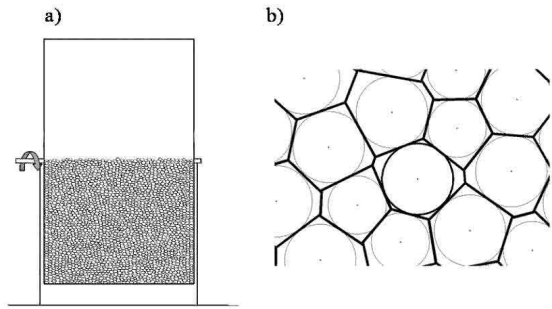}
\caption{(a) Experimental set-up. (b) Sketch of the modified Voronoi
tessellation for bidisperse systems.}
\label{flipflap}
\end{figure}

Let us now describe our experimental set-up, which is presented
schematically on figure~\ref{flipflap}(a). It is made of a bidimensional
rectangular glass container of width $50\, {\rm cm}$ and height
$60\, {\rm cm}$, with an inner gap of thickness $3.5\, {\rm mm}$. This
container is filled with a single layer of $5000$ cylindrical brass spacers
of two different sizes in equal number, resulting in a homogeneous and
disordered bidimensional packing.
The spacers (that we shall also call ``grains'') occupy approximately half
of the total volume of the container, as shown on figure~\ref{flipflap}(a).
The two kinds of spacers have a thickness of $3\, {\rm mm}$, and differ by
their diameter, which is respectively $d_s =4\, {\rm mm}$ for the small
spacers, and $d_l = 5\, {\rm mm}$ for the large ones.
For commodity, we choose $d_s$ as the unit length, so that the respective
areas of small and large disks are $v_s^0=\pi/4\simeq 0.78$ and
$v_l^0=\pi/4\,(d_l/d_s)^{2}\simeq 1.23$.

In addition to the two different kinds of grains which differ by their size,
we also use two different families of grains (each with small and large
grains) with very different roughness properties. The first family consists
in metal grains with a smooth surface, the static friction coefficient of which is approximately $\mu \approx 0.35$, whereas the second one is composed of
serrated rollers with a crenel shape on their peripheral surface, resulting in a static friction coefficient $\mu \approx 0.5$. Moreover different densities of the packing are obtained by vibrating it with a
hammer-like device installed below the container. 

The aim of the experiment is to generate a large number of different random 
configurations of the packing, and to analyze these configurations
through an image processing. The protocole is chosen in order to achieve a maximal randomness in the exploration of all the accessible configurations even though it is difficult to have a proper characterization of such a randomness. However we could verify that the individual distribution of voids and their repartition in space is very different from one image to another. An experimental run consists in $5000$ cycles defined by the following procedure. Starting from a vertical position, the cell is slowly rotated of $360^{\circ}$, at a constant angular velocity of $6^{\circ}/s$. During such a cycle, the grains fall from one side to the other and then back to the initial side. We let the packing relax for a few seconds so that it reach a mechanically stable state and
we take a picture of a suitably defined window within the bulk of the packing,
which contains approximately $1000$ grains. Once this is done, the whole
system is vibrated for $10\, {\rm s}$ at a frequency of $10\, {\rm Hz}$ and
with an amplitude of $5\, {\rm mm}$, to allow for rearrangements and
compaction. A second picture is then taken. Note that vibrating the system
does not aim at studying the compaction dynamics itself but simply at
generating a second packing state with a different density. 

The image processing consists in the following steps. The first step
is to locate in each picture the centers of the grains. Then a
Laguerre tessellation is used to associate to each grain a specific
area. This tessellation is a modified version of the Voronoi one with respect to
standard definitions, in order to take into account the bidispersity of the
grains~\cite{Okabe00}. An example of such a tessellation is shown on figure~\ref{flipflap}(b).
Once all cells are defined, the areas of the cells are computed and recorded,
keeping also track of the position and type (small or large) of the
associated grains.

Using the above procedure, we obtained three different sets of data,
corresponding respectively to smooth grains without vibration,
smooth grains with vibration, and rough grains without vibration.

\section{Experimental results}

\subsection{One-grain volume and free volume distributions}

\begin{figure}[b]
\hfill\includegraphics[width=6cm,height=4.5cm,clip]{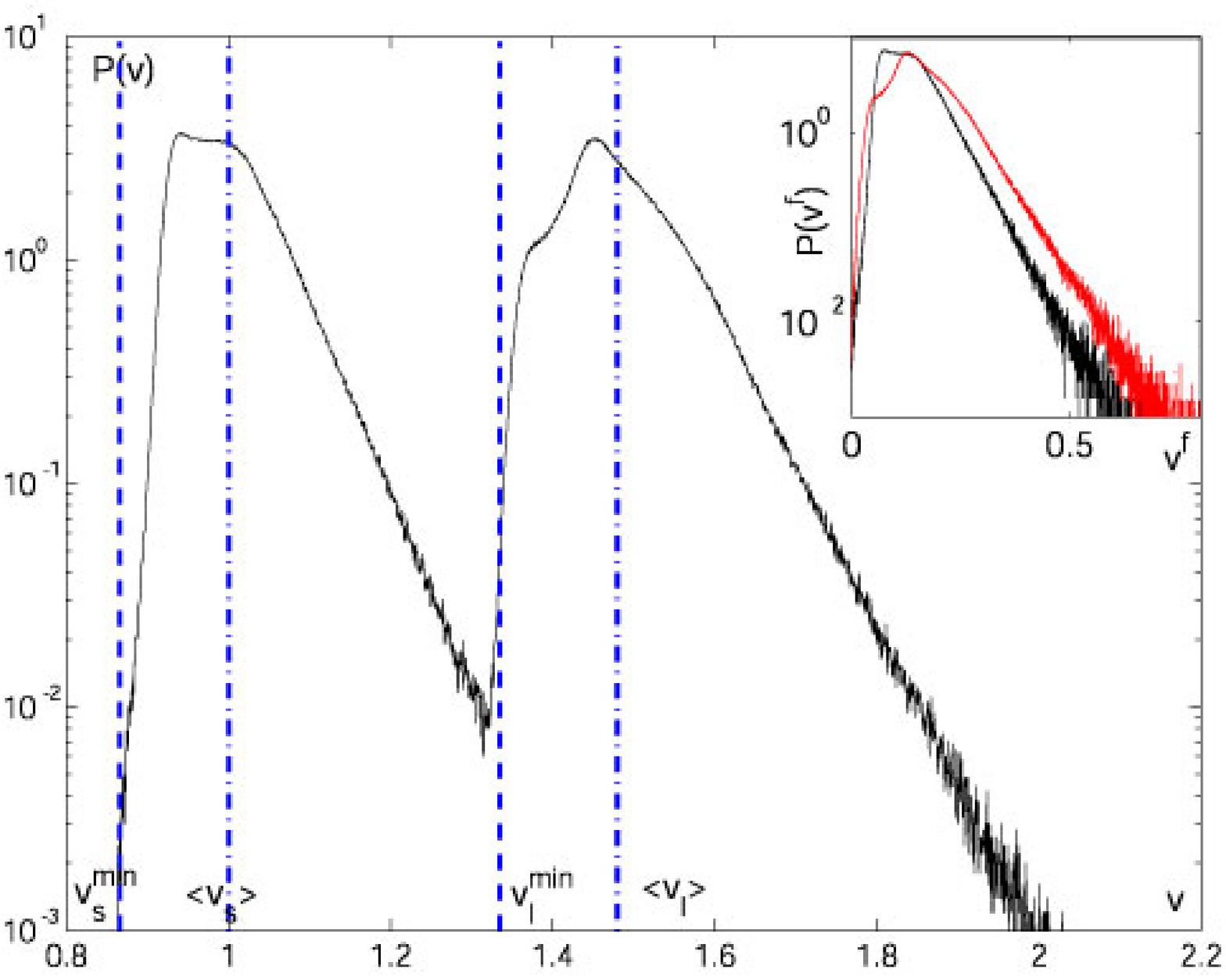}
\includegraphics[width=6cm,height=4.5cm,clip]{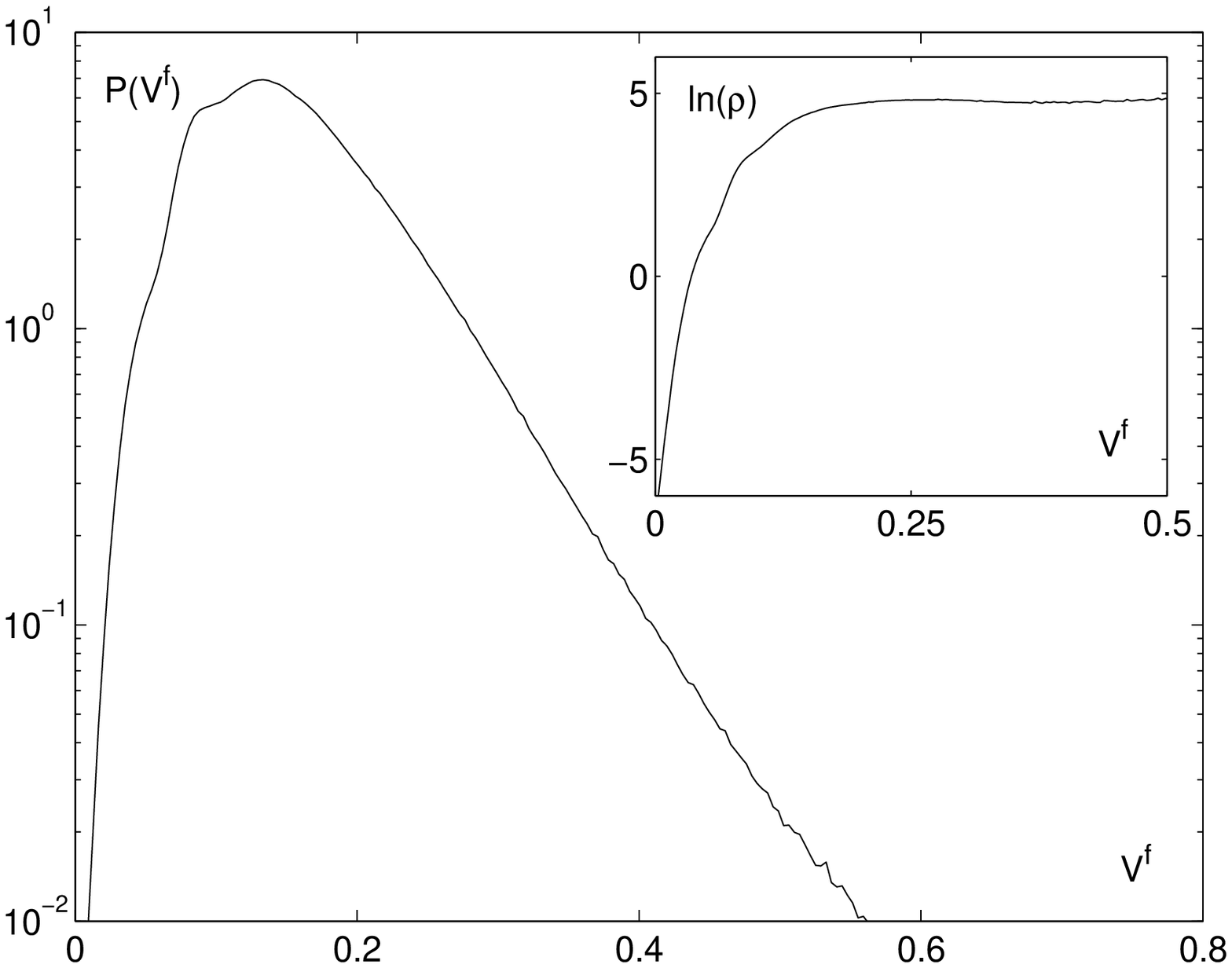}\\
\vbox{\hspace{6.5cm} (a) \hspace{5cm} (b)}
\caption{(a) Distribution of the Voronoi cells area for the non-vibrated
smooth grains. Vertical dashed lines : minimal Voronoi cell area. Vertical
dash-dotted lines : conditional average Voronoi cell area.
Inset: distributions of the free volume conditioned by the grain size;
(dark): small grains; (grey): large grains. (b) Distribution of the free
volume for the non-vibrated smooth grains including both sizes of grains.
Inset: density of states of one grain surrounded by a free volume $v^f$.}
\label{pdf_v_vf}
\end{figure}

Let us first concentrate on the non-vibrated smooth grains.
The distribution of the Voronoi cell areas defined above is presented
on figure~\ref{pdf_v_vf}(a).
This distribution displays two peaks centred on $<v_s> = 1.00$ and
$<v_l> = 1.49$, that corresponds respectively to the average area
occupied by the small and large grains, computed independently.
Also indicated on the figure are the minimal values that a Voronoi cell
can possibly take - the closest regular hexagon- for each type of grain,
$v_s^{min}=\sqrt{3}/2 \simeq 0.866$ and $v_l^{min}=\sqrt{3}/2(d_l/d_s)^2
\simeq 1.35$. Both peaks present a well-defined exponential tail, which is
easily isolated when considering the distributions of the free volume 
($v_{s,l}^f=v-v_{s,l}^{min}$), for each type of grain as shown on the inset
of figure~\ref{pdf_v_vf}(a). It shall be noticed that the characteristic free
volume for each type of grain, ${v_s^f}^*= 0.055$ and ${v_l^f}^* = 0.060$
are closer to each other than one would have expected given the particle
size ratio. In other words, the characteristic free volume accessible to
one grain not only depends on the size of that grain, but also on its
neighborhood, most often composed of grains of both sizes.

Figure~\ref{pdf_v_vf}(b) displays the distribution of the free volume
without conditioning it to the type of grains.  One still observes the
well-defined exponential tail, with a characteristic free volume
${v^f}^*=0.058$. In the spirit of Edwards' description, one may assimilate
this volume, characteristic of the sample, to a compactivity. This is
of course a rather strong interpretation of the data, since it would be
necessary at least to check that the dependence of the distributions
on this compactivity is indeed fully embedded in the exponential tail.
Still, as an indication, we have plotted in the inset of
figure~\ref{pdf_v_vf}(b) the quantity
$\ln(\rho(v^f))=\ln(P(v^f)))+v^f/{v^f}^*$, the density of states of one
grain surrounded by a free volume $v^f$. It saturates for large free volumes,
after a sharp increase for small values.

\begin{figure}[t]
\centering\includegraphics[width=8cm,height=5.5cm,clip]{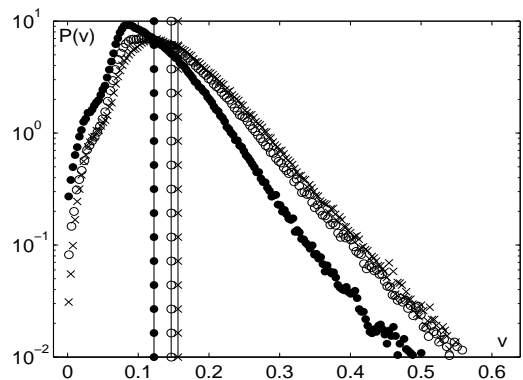}
\caption{Free volume distributions for the three different experimental runs:  non-vibrated smooth ($\circ$) and rough ($\times$) grains, and vibrated smooth grains ($\bullet$) (the vertical lines indicate the mean values).}
\label{pdf_vf1}
\end{figure}

We now characterize the dependence of the above free volume distribution
on both the roughness of the grains and the history of the packing.
Figure~\ref{pdf_vf1}  displays the free volume distributions for the
non-vibrated packings of smooth and rough grains, as well as for vibrated
and non-vibrated packings of smooth grains. The values of the mean and 
characteristic free volumes are reported in Table~\ref{tab:table1}. 
Obviously, the mean free volume decreases when vibrating the pile. It is
also lower in a pile of smooth grains than in a pile of rough grains prepared 
the same way: roughness allows more efficient arch building, hence larger
interstitial space. The characteristic free volume associated with the
exponential tail is smaller for the vibrated, more compact system,
as suggested in~\cite{Richard03,Aste05b}. Inside non-vibrated packings,
this characteristic free volume is not much sensitive to the roughness
of the grains.

\subsection{Volume and free volume distributions for clusters of neighboring
grains}

As explained in the introduction, our main goal in the present study is to
test experimentally the coherence of a statistical construction for
granular media. To do so, one has to study not only one-grain quantities,
but also the statistical properties of a subsystem made of a sufficiently
large number of grains. In addition, it is clear that a richer information
may be gained from the \emph{distributions} of global quantities
characterizing the system, rather than from simple averages of these
quantities. Indeed, distributions of global quantities have proved to be
a useful tool in the study of non-equilibrium systems, both from
the experimental~\cite{Bramwell98,Pinton99} and from the numerical
\cite{Bramwell00,Aumaitre01} point of view.
Moreover, it should be noticed that in granular media, one may not
expect the N-body probability distribution for the volumes of a set of $N$
grains to factorize as a product of one-grain distributions, due to
possible correlations between grains. Accordingly, some interesting and
non trivial properties may emerge from the statistical properties of
systems consisting of $N \gg 1$ grains, so that there is a strong motivation
to determine such statistical properties experimentally.

To this aim, we first define the system on which measurements are to be
performed. As the packing is disordered, there is no obvious way to
determine the system by a simple geometric boundary, contrary to what may
be done if grains were sitting on a regular lattice. To overcome this
difficulty, we define clusters of $N$ grains by starting from an arbitrary
grain, and looking for its $N-1$ nearest neighbors, that is, the $N-1$ grains
that are the closest to the original grain. The cluster is then composed
of the original grain and of the $N-1$ nearest neighbors.

\begin{figure}[b]
\hfill\includegraphics[width=6cm,height=4.6cm,clip]{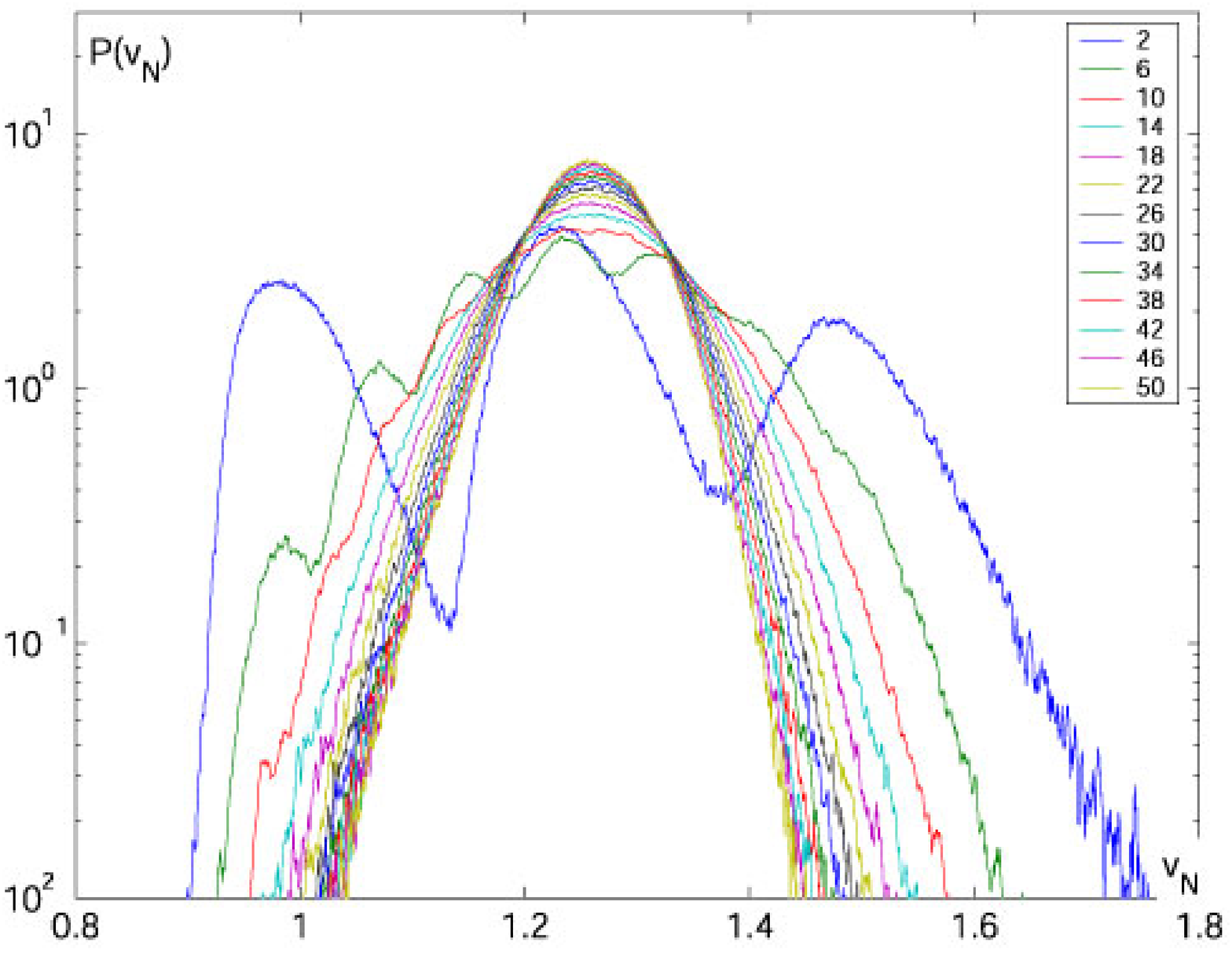}
\includegraphics[width=6cm,height=4.6cm,clip]{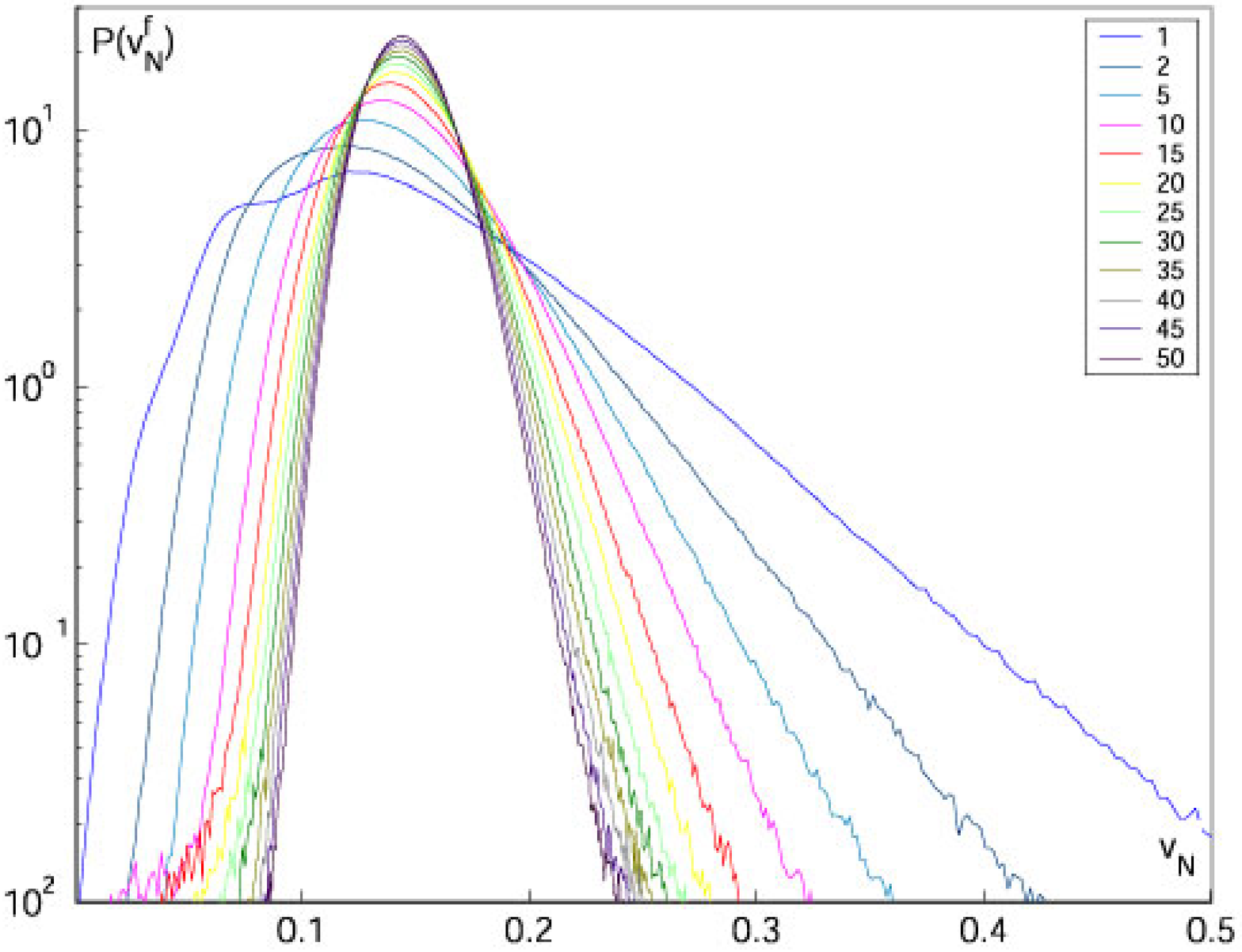}\\
\vbox{\hspace{6.5cm} (a) \hspace{5cm} (b)}
\caption{Distribution of (a) the volume and (b) the free volume per grain
inside clusters of N grains: the larger N, the narrower the distribution}
\label{pdf_vN}
\end{figure}

Now that the clusters of grains are defined, one would like to characterize
their statistical properties. At this stage, it is important to note
that the free volume and the full volume are no longer equivalent,
since clusters are composed of both small and large grain.
Figure~\ref{pdf_vN}(a) and (b) respectively display the distribution of
the volume per grain $v_N=N^{-1}\Sigma_{i=1}^N v_i$, and of the free volume
per grain $v^f_N=N^{-1}\Sigma_{i=1}^N v^f_i$, inside clusters of
$N$ neighboring grains.

The distribution of the free volume per grain $v^f_N$ converges much faster
towards a well defined distribution than that of the volume. Indeed, it is
clear from figure~\ref{pdf_vN}(a), in the case of the volume distributions, 
as $N$ increases from 1 to 10, that combinatory effects induced by the
bidispersity give rise to series of peaks which complexify the distribution.
For $N=2$, one observes three peaks, which are related to the three
two-particle cluster configurations. Similarly, for $N=3$ --not shown here--,
the distribution exhibits four peaks, etc. Such effects are absent in the
case of the free volume distributions (figure~\ref{pdf_vN}b). Also,
describing the volume distribution imposes to know the minimal value of the
volume per grain, a non trivial quantity for clusters composed of grains of
both sizes, whereas by construction, the distribution of the free volume
readily addresses this issue. Accordingly, we restrict the following analysis
to the free volume distributions. 

\begin{figure}[t]
\hfill\includegraphics[width=6cm,height=4.5cm,clip]{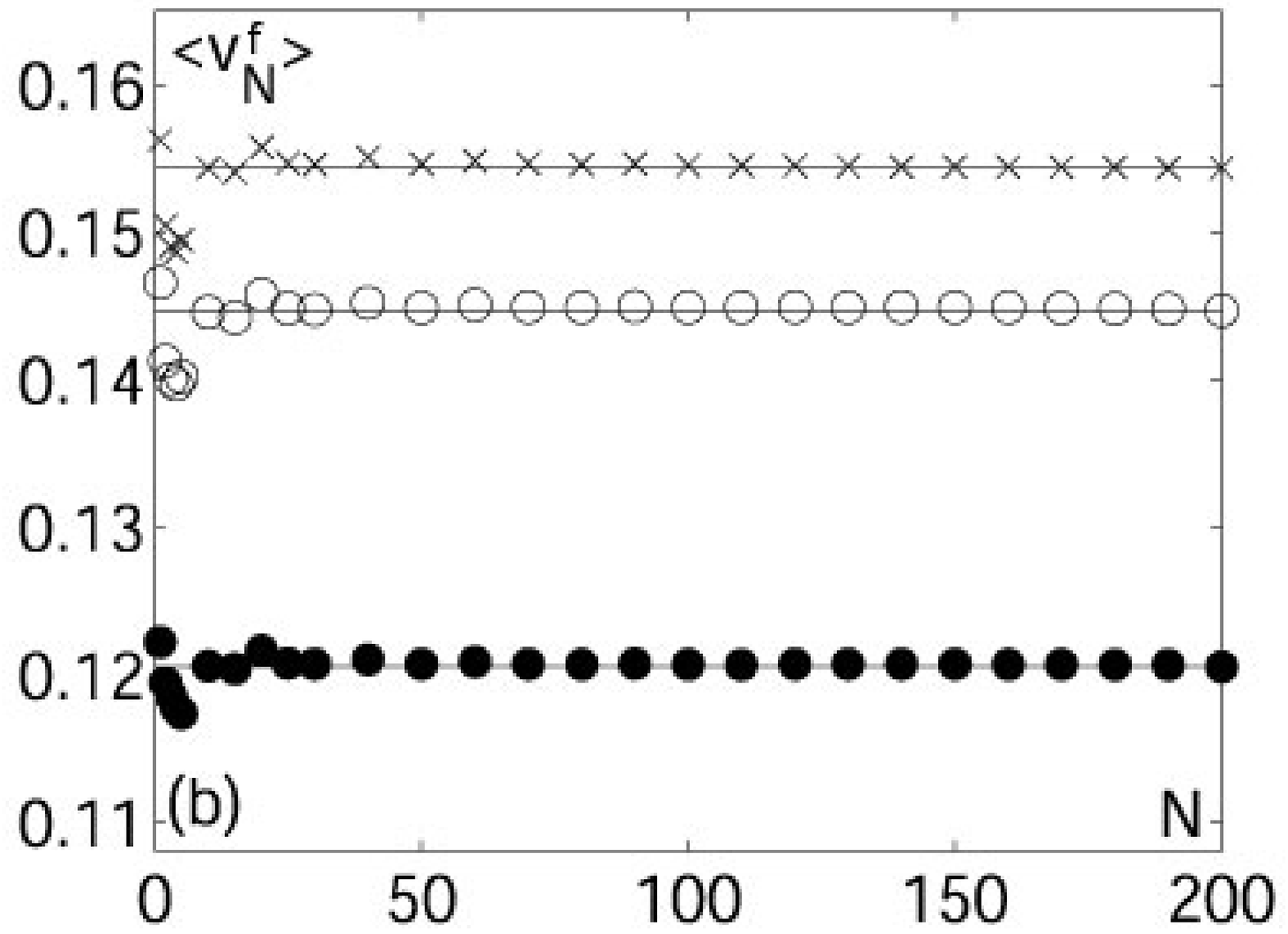}
\includegraphics[width=6cm,height=4.5cm,clip]{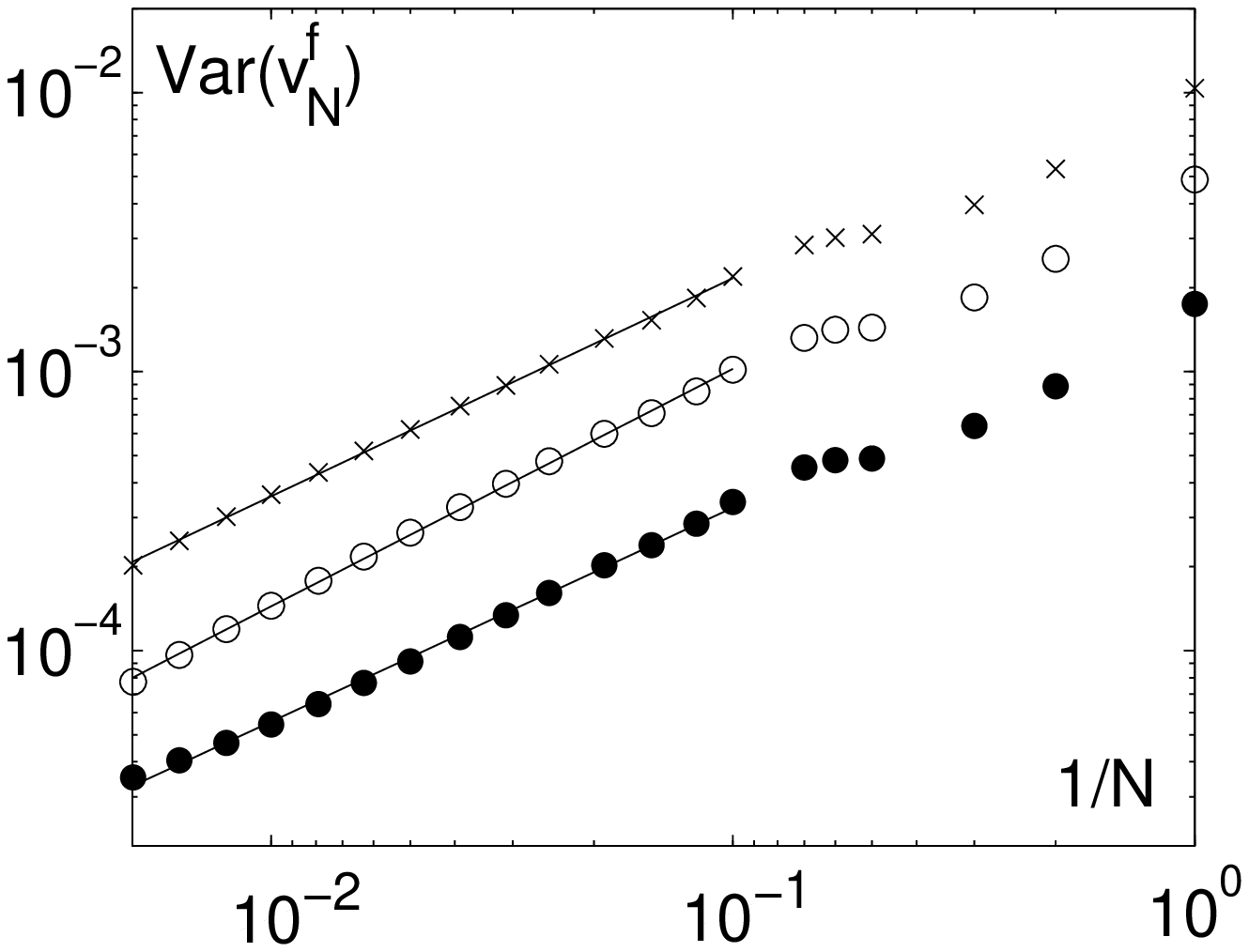}\\
\vbox{\hspace{6.5cm} (a) \hspace{5cm} (b)}
\caption{Dependence on $N$ of $\langle v^f_{N} \rangle$ (a) and
$\mathrm{Var}(v^f_N)$ (b) for the three different experimental runs
(same symbols as in Fig.~\ref{pdf_vf1}). Curves in (b) are shifted for
clarity.}
\label{pdf_moments}
\end{figure}

Given that the free volume per grain $v^f_N$ may take arbitrary positive
values, and that the distribution $P(v^f_N)$ seems to have an exponential
tail, at least for small $N$, it is natural to try to describe our data
with Gamma laws. This description is also supported by numerical simulations
of Poisson-Voronoi cells~\cite{Jarai04}.
We thus use the following ansatz
\be
P(v^f_N)=\frac{1}{\chi_N^{\eta_N}\Gamma(\eta_N)}(v^f_N)^{\eta_N-1}
e^{-v^f_N/\chi_N}
\label{gamma-dist}
\ee
which involves two unknown parameters $\eta_N$ and $\chi_N$; $\eta_N$ may be
considered as a shape parameter whereas $\chi_N$ is a scale parameter. Note
that at this stage, one shall not anticipate any relation between $\chi_N$
and the compactivity as defined by equation~(\ref{compactivity}).
A simple way of determining $\eta_N$ and $\chi_N$ is to adjust their values
so that the first two moments of the Gamma law coincide with the experimental
data. To this aim, we make use of the relations $\langle v^f_{N} \rangle =
\eta_N \chi_N$ and $\mathrm{Var}(v^f_N) = \eta_N \chi_N^2$, from which one
deduces $\chi_N = \textrm{Var}(v^f_N)/\langle v^f_{N} \rangle$ and
$\eta_N=\langle v^f_{N} \rangle/\chi_N$.
The average free volume $\langle v^f_{N} \rangle$ is plotted on
figure~\ref{pdf_moments}(a), and is seen to rapidly converge to a constant
value $v^f_{\infty}$, the mean free volume of the packing.
A more surprising behavior is obtained from the variance of $v^f_N$, when
plotted as a function of $1/N$, as shown on figure~\ref{pdf_moments}(b).
If the individual free volumes associated with each grains were independent
random variables, the variance of the total free volume of the cluster
would scale as $N$, so that free volume per grain in the cluster would
be proportional to $1/N$. On the contrary, for the range of cluster sizes
over which measures have been performed, the variance $\mathrm{Var}(v^f_N)$
scales as
\be
\mathrm{Var}(v^f_N) \approx \nu N^{-\alpha}, \quad 0<\alpha < 1
\ee
The above large $N$ behavior of the two first moments leads us to the
following scaling $\chi_N=\chi_e N^{-\alpha}$ and $\eta_N = \eta_e
N^{\alpha}$, with $\chi_e=\nu/v^f_{\infty}$ and $\eta_e={v^f_{\infty}}^2/\nu$.
The values of the different parameters are reported in Table~\ref{tab:table1}.

\begin{figure}[b]
\centering\includegraphics[width=8cm,height=5cm,clip]{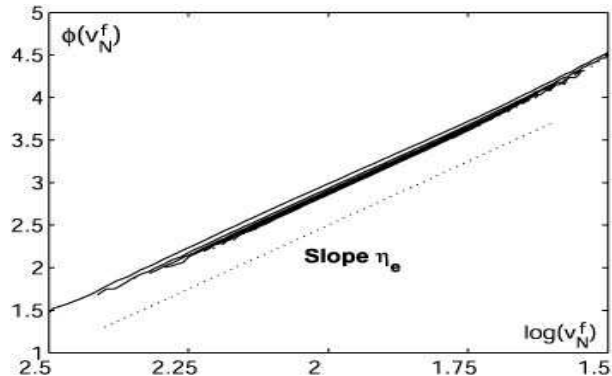}
\caption{\sl Scaling function $\Phi(v^f_N)$ as a function of $\ln v^f_N$, for $N=10,30,50...190$ for the non-vibrated smooth grains.}
\label{scaling}
\vspace{-3mm}
\end{figure}

Altogether, for large $N$ one can rewrite the distributions $P(v^f_N)$ in the
following form, making the scaling with $N$ explicit

\be \label{non-extensive-PvN}
P(v^f_N) = \frac{N}{Z_c(\chi_e,\eta_e ,N)}\exp\left[N^\alpha \left(\eta_e \ln
v^f_N - \frac{v^f_N}{\chi_e}\right)\right],
\ee
where the normalization factor $Z_c(\chi_e,\eta_e, N)$ is given by
\be
Z_c = \exp \left[N^\alpha \eta_e \left(\ln \frac{\chi_e}{\eta_e}-1\right)
\right].
\ee
Hence a non-extensive scaling appears over the explored range of cluster sizes
for $\ln P(v^f_N)$. Figure~\ref{scaling} displays the scaling function, $\Phi(v^f_N)=N^{-\alpha}\ln [P(v^f_N)\, \exp(N^{\alpha} v^f_N/\chi_e)]$. The collapse and the linearity of the data confirm the validity of the above analysis, and yields for the distribution $\Phi_N(V^f)$ of the free volume $V^f=Nv_N^f$ for a subsystem of $N$ grains :
\be
\Phi_N(V^f) = \frac{1}{Z_c(\chi_e,\eta_e ,N)}\exp\left[N^\alpha \eta_e \ln
\left(\frac{V^f}{N}\right) - \frac{V^f}{N^{1-\alpha} \chi_e}\right].
\ee

In the extensive case $(\alpha=1)$, one would recover the general form of
equation (\ref{V_dist}), hence a formal identification of $\chi_e$ with the
compactivity $X$. Here this is not the case and one does not have a
compactivity independent of the system size, when applying the definition of
equation (\ref{compactivity}). However, one can still define generalized
entropy and ``free energy'' densities $s(v^f_N)$ and $f(v^f_N)$ by
\be
s(v^f_N)=\eta_e \ln v^f_N \quad {\rm and} \quad
f(v^f_N)=v^f_N-\chi_e s(v^f_N)
\ee
which allows $P(v^f_N)$ to be rewritten as
\be
P(v^f_N)=\frac{N}{Z_c} e^{-N^{\alpha}f(v^f_N)/\chi_e}
\ee
a form similar to the equilibrium one, except for the exponent $\alpha \ne 1$.
With such definitions one recovers, from a saddle-point calculation on
equation~(\ref{non-extensive-PvN}), the relation $\chi_e^{-1} =
\partial s/\partial v^f \vert_{\langle v^f\rangle}$, which generalizes the
one usually used in equilibrium extensive statistical mechanics and suggest
that $\chi_e$ can still be interpreted as the compactivity of the packing.

It should be noticed that the above canonical analysis implicitely assumes that the total volume of the packing is constant, so that, when considering a cluster of $N$ grains, the rest of the packing plays the role of a reservoir. In the present finite system we can estimate the total volume fluctuations to be between $1.7\% $ and $0.8\% $ depending on whether we assume the sub-extensive scaling to hold up to the system size or not. This is consistent with a variation of half a layer of grain at the top of the pile.

\begin{table}[t]
\centering
\begin{tabular}{cccc}
\hline
&$\circ$ &$\bullet$ &$\times$\\
\hline
$v^f_{\infty}$& .145 & 0.120 & 0.154\\
${v^f}^*$& 0.059$\pm$0.001 & 0.046$\pm$0.001 & 0.061$\pm$0.001 \\
$\chi_e$& 0.048$\pm$0.0015 & 0.032$\pm$0.0015 & 0.040$\pm$0.0015\\
$\eta_e$& 3.0$\pm$0.1& 3.7$\pm$0.1 & 3.9$\pm$0.1\\
$\alpha$& 0.85$\pm$0.02& 0.80$\pm$0.02& 0.77$\pm$0.02\\
\hline
\end{tabular}
\caption{\label{tab:table1}\sl Summary of the different parameters extracted
from the distributions of free volume in the three systems
(same symbols as in Fig.~\ref{pdf_vf1}).}
\end{table}

\section{Discussion}

\subsection{Physical origin of the non-extensivity}

As explained above, our central result is the observation of a sub-extensive
scaling, with the cluster size $N$, of the variance of the free volume per
grain in clusters of increasing sizes. This unusual scaling qualitatively
indicates the presence of correlations between the individual free volumes
of the grains, as otherwise $\mathrm{Var}(v^f_N)$ would scale as $1/N$.

Let us try to discuss more quantitatively how the departure from the
$1/N$-behavior for $\mathrm{Var}(v^f_N)$ relates to the presence of
correlations in the system.
To this aim, we make the following simplifying assumptions:
(i) the cluster has the shape of a disk, with a radius $R$;
(ii) the density of grains is homogeneous inside the cluster;
(iii) introducing $\delta v_i^f \equiv v_i^f - \overline{v}^f$,
where $\overline{v}^f$ is the average free volume per grain,
the correlation $\langle \delta v_i^f \delta v_j^f \rangle$ depends
only on the relative distance $r=|\mathbf{r}_i - \mathbf{r}_j|$, and not
on further information like the position of the grains $i$ and $j$ with
respect to the center of the cluster. This correlation is denoted as $C(r)$
in what follows. Expanding the variance, one finds
\be
\mathrm{Var}(v^f_N) = \mathrm{Var}\left(\frac{1}{N}\sum_i v_i^f\right)
= \frac{1}{N^2} \sum_{i,j} \langle \delta v_i^f \delta v_j^f \rangle
\ee
Under the above assumptions, one can then write
\be \label{var-correl}
\mathrm{Var}\left(\frac{1}{N}\sum_i v_i^f\right) = 
\left(\frac{\rho}{N}\right)^2 \int_{|\mathbf{r}_1|<R}
d\mathbf{r}_1 \int_{|\mathbf{r}_2|<R} d\mathbf{r}_2 \,
C(|\mathbf{r}_1 - \mathbf{r}_2|)
\ee
Let us now assume that the correlation function $C(r)$ decays like
$r^{-\gamma}$ for large $r$. Then the large $R$ behavior of the integral
given in Eq.~(\ref{var-correl}) can be computed, and turns out to be 
$\mathrm{Var}(v^f_N) \sim N^{-\gamma/2}$. To obtain this result, one may use
polar coordinates to integrate over the vectors $\mathbf{r}_1$ and
$\mathbf{r}_2$. The angular part of the integral actually appears only in
the prefactor of the power law $N^{-\gamma/2}$. This suggests that the
precise shape of the cluster only plays a minor role, and does not affect
the scaling behavior itself. In the present case, given the values of
$\alpha$ reported above, one would infer that correlations decay like
$r^{-\gamma}$ with $\gamma \approx 1.6$ over the experimentally accessible
range of distances.

\begin{figure}[b]
\hfill\includegraphics[width=6cm,height=4.5cm,clip]{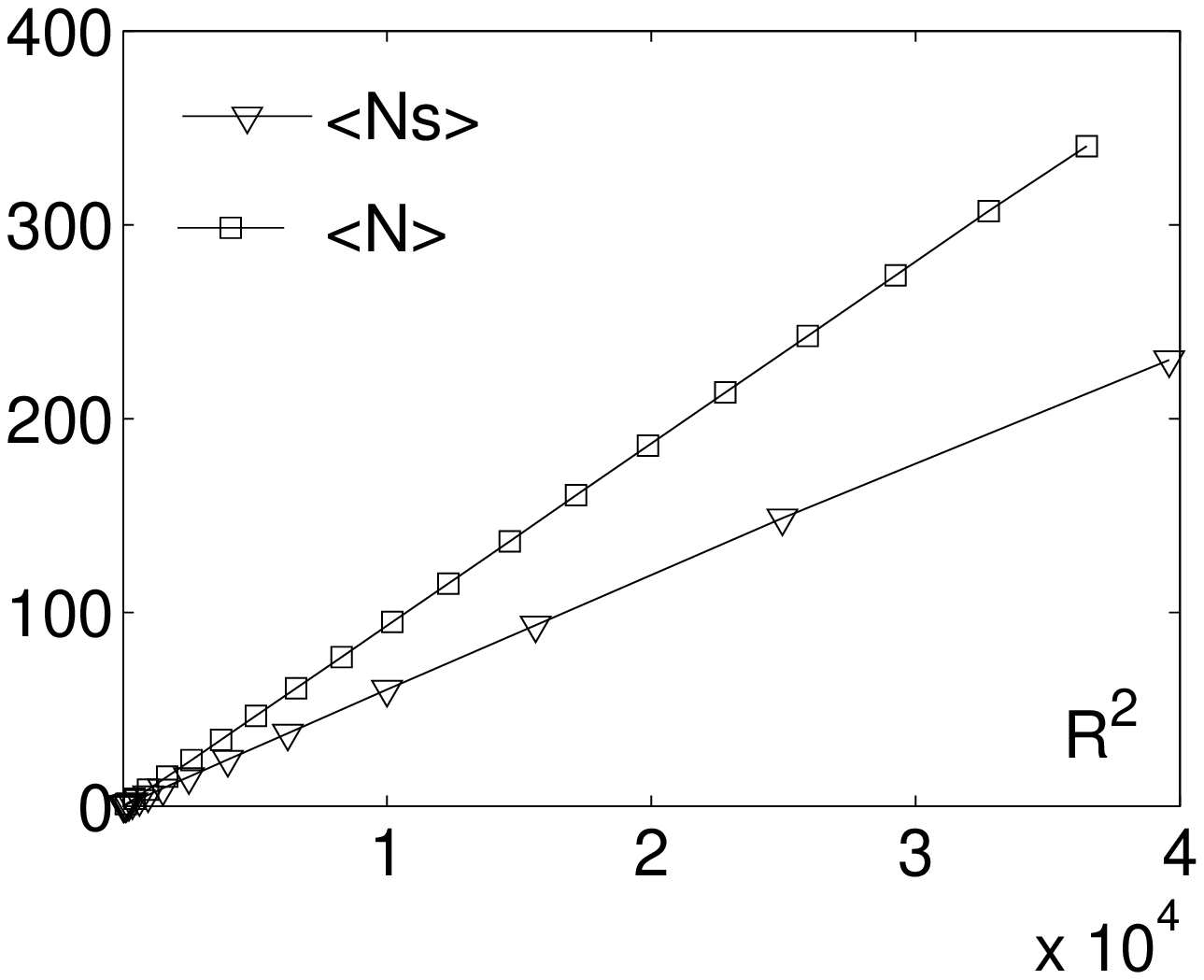}
\includegraphics[width=6cm,height=4.5cm,clip]{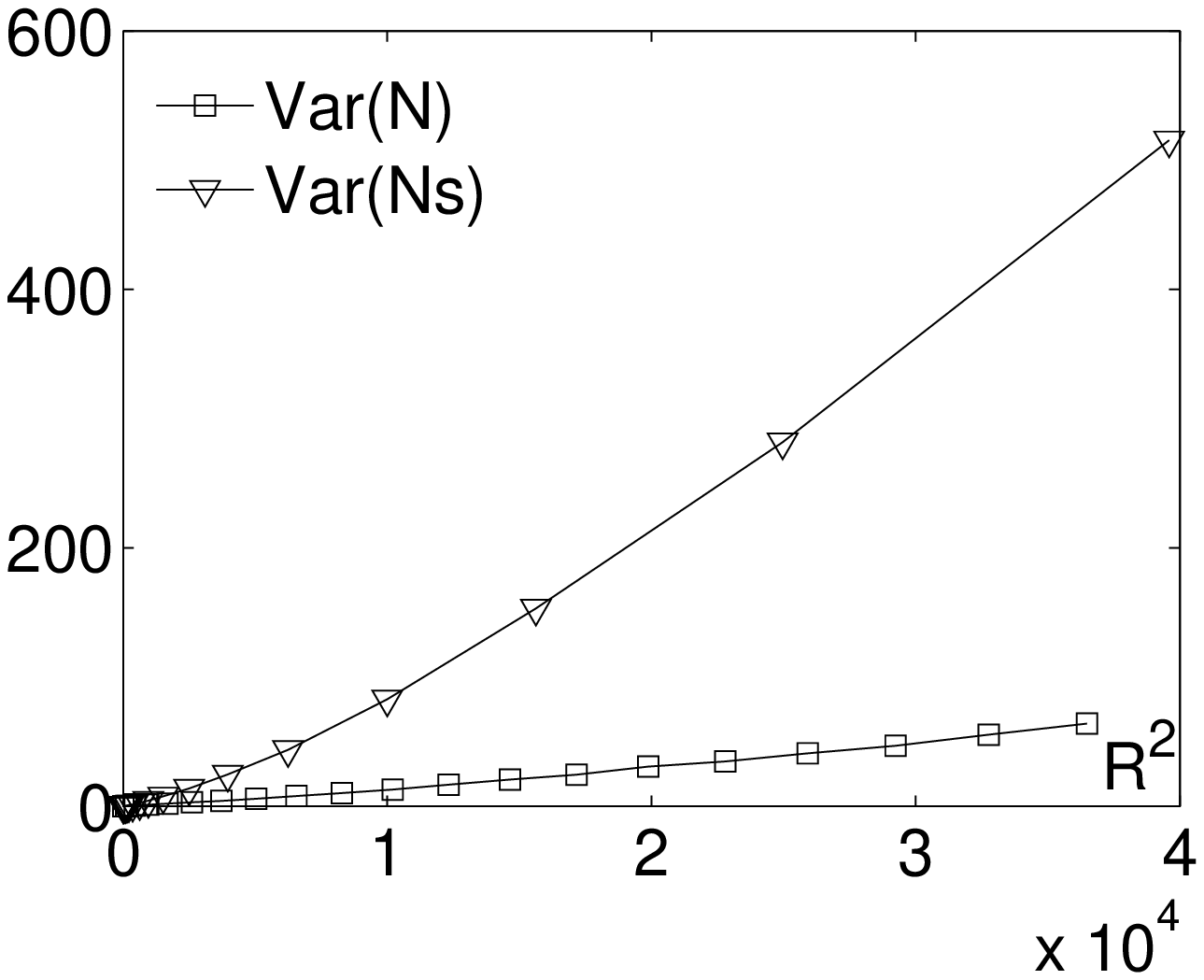}\\
\vbox{\hspace{6.5cm} (a) \hspace{5cm} (b)}
\caption{First (a) and second (b) moments of the number of grains
$\langle N \rangle$ and small grains $\langle N_s \rangle$ inside a disk of
radius $R$.}
\label{moy-var-densite}
\end{figure}

From a physical point of view, such correlations seem to be related to
anomalous fluctuations of the numbers of particles inside the clusters.
More precisely, one may think either about anomalous fluctuations of the
density of particles~\cite{Donev05}, or to fluctuations of the fraction of 
small and large particles inside a cluster.  Figure~\ref{moy-var-densite}(a)
displays the averaged number of grains $\langle N \rangle$ and the averaged
number of small grains $\langle N_s \rangle$ inside a disk of radius $R$
(note that this is a procedure different from the one previously used,
which instead involved clusters of $N$ neighboring grains).
As expected both scale like $R^2$ confirming the visual observation that the
granular packing is compact.
Conversely, $\mathrm{Var}(N)$ and $\mathrm{Var}(N_s)$, shown on
figure~\ref{moy-var-densite}(b), do not scale like $R^2$, and hence not
like $N$ or $N_s$ respectively. Note that such anomalous fluctuations are
much stronger for the fraction of small particles than for the density itself.
The fact that the average value scale precisely like $R^2$ is interesting,
as it shows that trivial finite-size corrections due to, e.g., the
discrete character of granular media, are very small in the range of
sizes explored. This confirms that the anomalous scaling of the variance 
is relevant, and is not simply due to generic finite-size corrections
affecting any observable.

Such anomalous fluctuations and scalings could be originated in the two-dimensional confinement of our system, and one may wonder whether the same exponent would be expected in three dimensions. Such experiments have not been conducted in three dimensions. However one element of answer goes as follows. We ran simulations of Poissonian Voronoi tesselations in two dimensions (that is, with point-like particles thrown at random on a plane), performed the same size scaling analysis and did not observe any anomalous scaling. This underlines the role of the finite extension of the grains in the observed scalings.

\subsection{Interpretation of the parameters $\eta_e$ and $\chi_e$}

Interestingly, even if extensivity does not hold, the above experimental
study still leads to a consistent statistical description.
In particular, the relation
$\chi_e^{-1} = \partial s/\partial v^f \vert_{\langle v^f\rangle}$,
which generalizes the one usually used in equilibrium extensive statistical
mechanics, may be recovered from a saddle-point calculation on
equation~(\ref{non-extensive-PvN}).
This results indeed gives support to the interpretation of $\chi_e$ as the
compactivity of the packing. It is also consistent with the definition
proposed for thermodynamic parameters like temperature in equilibrium
systems with long range interactions~\cite{Curilef99}. In this case,
these parameters are essentially rescaled (with respect to more usual
definitions) by a size-dependent effective number of degrees of freedom, 
so that they remain finite in the thermodynamic limit.
We also note that the qualitative behavior of $\chi_e$ is the one physically
expected. For instance, when lowering the volume fraction by vibrating the
packing, $\chi_e$ is seen to decrease, in a way similar to what happens
in an equilibrium system, where temperature decreases when lowering energy.

The second parameter describing the statistics of the packing, $\eta_e$,
rather characterizes the entropy $s(v^f_n)$. In more physical terms,
$\eta_e N^{\alpha}$ may be thought of as the effective number of independent
(collective) objects in a cluster of $N$ grains.
This may be seen from the most simple case of the Poisson-Voronoi cells
in one dimension~\cite{Jarai04}. In this case, points are thrown at
random on a line, and the length of the associated Voronoi cells is given
by half the sum of the length of the two segments surrounding the point
considered. As the lengths of the segments between two adjacent points
are independent random variables, a Voronoi cell may be thought of as
having two internal degrees of freedom, thus suggesting $\eta=2$.
This simple reasoning is confirmed by exact computations~\cite{Jarai04}.
So coming back to our experimental situation, this suggests that the presence
of correlations leads to complex and effectively independent objects, with
possibly a fractal shape, the number of which is growing non-extensively with
the size of the cluster.

\subsection{Consequences for the thermodynamic formalism}

The above results have important consequences from the point of view of
measurements of statistical properties in granular systems.
On the one hand, they show that it is important to check the extensivity
of the system when extracting the compactivity from usual thermodynamic
relations. The non-extensive scaling behavior is expected to hold up to
a scale comparable to the correlation length of the packing.
It has been recently proposed that amorphous media may be in a state
close to a critical state associated with the ``jamming transition''
\cite{Hern02,Hern03,Wyart05}. Hence, this suggests that the correlation
length may be quite large in disordered granular packings.

Other consequences also need to be stressed. The parameter $\eta_e$ appearing in the expression of the entropy varies together with $\chi_e$ when the way to prepare the system is changed. This is a signature of the non-equilibrium character of the system. In an equilibrium situation, changing the preparation conditions would amount to change, say, the energy and the temperature of the system, but it would not affect the functional expression of the entropy. Only the value of the entropy would change. Here the value of $\eta_e$ is determined by the microscopic dynamics and the change in $\eta_e$ suggests that the way microscopic configurations are sampled by the preparation process changes drastically when varying the density: not only the typical volumes sampled are modified, but also the way configurations of a given volume are explored.

It is worth underlining that such a situation could formally occur in the case of Edwards' approach. Indeed, since Edwards' equiprobability assumption is restricted to the states blocked under the considered dynamics, the density of such states may a priori depend on that dynamics. From that point of view, the generalized formalism proposed in Sect.~2, takes into account any possible departure from the equiprobability of microstates, including the specific case of a dynamical selection of the blocked states. The drawback is then that the probability weights $f(\mathcal{C})$ are unknown, but the relevant macroscopic information may be encoded into the parameter $\eta_e$, which is experimentally accessible.

Finally, we note that the presence of correlations ``renormalizes'' in some
way the ``compactivity'' $\chi_e$, so that it is no longer equal to the
characteristic volume $v^{f*}$ of the exponential decay of the one-grain
distribution. Although one expects that extensivity is recovered for system
sizes larger than a crossover length scale related to the correlation length,
there is no reason to believe that this ``large scale compactivity'' would
be equal to $v^{f*}$.
Hence, this shows that measuring the slope of the exponential tail of the
one-grain distribution is not a relevant determination of compactivity.

\section{Conclusions}
In this paper, we have studied clusters of increasing sizes in a bidimensional
granular packing. Our central result is that a sub-extensive scaling of
the fluctuations of free volume appears. This strongly suggests the presence
of correlations between the individual free volumes of distant grains,
over the range of sizes explored experimentally.
In spite of this nonextensive scaling which partly rules out the
statistical formalism proposed for granular materials, it is still possible
to define in a consistent way an intensive compactivity.
This parameter is found to differ significantly from the one that may be
inferred from the exponential tail of the one-grain distribution.
In some sense, this leads to the idea that correlations act as a
``renormalization'' of the statistical parameters.
>From a more general point of view, it also shows that one has to be careful
when applying statistical concepts to a system for which the microscopic
dynamics is poorly understood --namely a kind of ``black box''.
In order to be sure that the procedure is well enough controlled, one
should make some checks of the fundamental properties of the system,
among which the extensivity plays a major role.

\ack
We are grateful to all the participants to our ``Glassy Working Group''
in Saclay and in particular J.-P. Bouchaud for enriching and enlightening
discussions. We thank C. Gasquet and V. Padilla for technical assistance.

\section*{References}

\bibliography{bibGlass}

\end{document}